\documentclass[conference]{IEEEtran}
\IEEEoverridecommandlockouts
\usepackage{cite}
\usepackage{amsmath,amssymb,amsfonts}
\usepackage{algorithmic}
\usepackage{graphicx}
\usepackage{textcomp}
\usepackage[table]{xcolor}
\usepackage{tikz}
\usetikzlibrary{patterns}
\usepackage{pgfplots}
\usepackage{pgf-pie}
\usepackage{pgfplotstable}
\usepackage{booktabs}
\usepackage{enumitem} 
\usepgfplotslibrary{colormaps}
\usepackage{colortbl} 
\pgfplotsset{compat=newest}
\usepackage{tabularx}
\usepackage{csquotes}

\usepackage{caption}
\usepackage{subcaption}

\pgfplotsset{
    discard if/.style 2 args={
        x filter/.code={
            \edef\tempa{\thisrow{#1}}
            \edef\tempb{#2}
            \ifx\tempa\tempb
                \def\pgfmathresult{inf}
            \fi
        }
    },
    discard if not/.style 2 args={
        x filter/.code={
            \edef\tempa{\thisrow{#1}}
            \edef\tempb{#2}
            \ifx\tempa\tempb
            \else
                \def\pgfmathresult{inf}
            \fi
        }
    }
}

\usepackage{listings}
\newcommand{\elmi}[1]{\lstinline[language=haskell]{#1}}

\usepackage{pifont}
\newcommand{\xmark}{\ding{55}}%

\definecolor{INTERpos}{RGB}{188, 236, 172} 
\definecolor{ELMpos}{RGB}{178, 255, 102} 
\definecolor{ONL}{RGB}{178, 229, 102} 
\definecolor{IRL}{RGB}{229, 229, 102} 
\definecolor{ANX}{RGB}{255, 204, 102} 
\definecolor{CODE}{RGB}{255, 178, 102} 
\definecolor{INTERneg}{RGB}{255, 102, 102} 
\definecolor{FOCUS}{RGB}{0, 119, 179} 
\definecolor{CAPT}{RGB}{255, 153, 255} 
\definecolor{GPT}{RGB}{204, 153, 255} 
\definecolor{DEEPEN}{RGB}{85, 0, 204} 
\definecolor{QAmov}{RGB}{153, 153, 255} 
\definecolor{NOISE}{RGB}{102, 178, 255} 
\definecolor{LENinc}{RGB}{102, 204, 255} 
\definecolor{LENdec}{RGB}{102, 229, 204} 
\definecolor{ROUNDpos}{RGB}{153, 204, 0} 
\definecolor{TIME}{RGB}{197, 204, 102} 
\definecolor{DEBAT}{RGB}{204, 204, 102} 
\definecolor{BREAK}{RGB}{204, 179, 102} 
\definecolor{ROUNDneg}{RGB}{150, 54, 86} 
\definecolor{ELMneg}{RGB}{229, 102, 102} 
\definecolor{RUBR}{RGB}{204, 102, 153} 
\definecolor{OUTCOME}{RGB}{204, 102, 204} 
\definecolor{PROCESS}{RGB}{153, 102, 204} 
\definecolor{PROF}{RGB}{102, 102, 204} 
\definecolor{LIVE}{RGB}{102, 153, 204} 
\definecolor{ACCOUNT}{RGB}{102, 178, 204} 
\definecolor{PACE}{RGB}{102, 204, 204} 
\definecolor{ANON}{RGB}{102, 204, 178} 
\definecolor{SIZE}{RGB}{210, 210, 162} 
\definecolor{PERS}{RGB}{255, 255, 178} 
\definecolor{PREREQ}{RGB}{255, 204, 178} 
\definecolor{RESOURCE}{RGB}{255, 204, 255} 
\definecolor{ENGL}{RGB}{178, 178, 255} 
\definecolor{TIMEMGMT}{RGB}{204, 153, 201} 
\definecolor{QAinc}{RGB}{224, 224, 255} 
\definecolor{CONCISE}{RGB}{224, 255, 255} 
\definecolor{RELEVANT}{RGB}{224, 224, 224} 
\definecolor{COMMON}{RGB}{178, 178, 178} 

\makeatletter
\def\pgfpie@slice#1#2#3#4#5#6#7#8{%
  \pgfmathparse{0.5*(#1)+0.5*(#2)}
  \let\pgfpie@midangle\pgfmathresult

  \path (#8) -- ++({\pgfpie@midangle}:{#5}) coordinate (pgfpie@O);

  \pgfmathparse{(#7)+(#5)}
  \let\pgfpie@radius\pgfmathresult
  
  \draw[line join=round,fill={#6},\pgfpie@style] (pgfpie@O) -- ++({#1}:{#7}) arc ({#1}:{#2}:{#7}) -- cycle;

  \pgfpie@ifchangedirection{%
    \pgfmathparse{min(((#1)-(#2)-10)/110*(-0.3),0)}
  }{%
    \pgfmathparse{min(((#2)-(#1)-10)/110*(-0.3),0)}
  }%
  
  \pgfmathparse{(max(\pgfmathresult,-0.5) + 0.8)*(#7)}
  \let\pgfpie@innerpos\pgfmathresult

  \pgfpie@ifx\pgfpie@text\pgfpie@text@inside{%
    \path (pgfpie@O) -- ++({\pgfpie@midangle}:{\pgfpie@innerpos}) node[align=center]
    {\pgfpie@scalefont{#3}\pgfpie@labeltext{#4}\\\pgfpie@numbertext{#3}};
  }{%
    \pgfpie@ifhidelabel{}{%
      \pgfpie@iflegend{}{%
        \path (pgfpie@O) -- ++ ({\pgfpie@midangle}:{\pgfpie@radius})
        node[inner sep=0, \pgfpie@text={\pgfpie@midangle:#4}]{};
      }%
    }%

    \ifnum#3>0
      \path (pgfpie@O) -- ++({\pgfpie@midangle}:{\pgfpie@innerpos}) node
      {\pgfpie@scalefont{#3}\pgfpie@numbertext{#3}};
    \fi
  }%
}
\makeatother

\def\BibTeX{{\rm B\kern-.05em{\sc i\kern-.025em b}\kern-.08em
    T\kern-.1667em\lower.7ex\hbox{E}\kern-.125emX}}
\begin{document}

\title{A Problem-Based Learning Approach to Teaching Design in CS1}

\author{
\IEEEauthorblockN{Christopher William Schankula, Habib Ghaffari Hadigheh,}
\IEEEauthorblockN{Spencer Smith, \& Christopher Kumar Anand}
\IEEEauthorblockA{\textit{McMaster University} \\
\textit{Department of Computing and Software}\\Hamilton, Canada \\ \{schankuc,ghaffh1,smiths,anandc\}@mcmaster.ca}}

\maketitle

\begin{abstract}
    Design skills are increasingly recognized as a core competency for software professionals. Unfortunately, these skills are difficult to teach because design requires freedom and open-ended thinking, but new designers require a structured process to keep them from being overwhelmed by possibilities. We scaffolded this by creating worksheets for every Design Thinking step, and embedding them in a PowerPoint deck on which students can collaborate. We present our experience teaching a team design project course to 200 first-year-university students, taking them from user interviews to functional prototypes. To challenge and support every student in a class where high school programming experience ranged from zero hours to three computer science courses, we gave teams the option of developing single-user or multi-user (distributed) web applications, using two Event-Driven Programming frameworks. We identified common failure modes from previous years, and developed the scaffolded approach and problem definition to avoid them. The techniques developed include using a ``game matrix'' for structured brainstorming and developing projects that require students to empathize with users very different from themselves. We present quantitative and qualitative evidence from surveys and focus groups that show how these strategies impacted learning, and the extent to which students' awareness of the strategies led to the development of metacognitive abilities.
\end{abstract}

\section{Introduction}
\label{s:1}

Design is an open-ended endeavour aimed at creating
novel solutions to novel problems, and yet new practitioners need a structured process to keep 
them from being overwhelmed and running into common failure modes. 
Designerly ways of thinking is as much a mindset as a skillset,
and starting young makes this easier to acquire, while also requiring stronger scaffolding.
We have found Design Thinking (DT) to be a well-structured approach to Human-Centred Design (HCD) for beginners.  
It is used in multiple disciplines, demonstrating generalizability and flexibility.
In this experience report,
we describe our first-year ``Introduction to Software Design'' project course
designed for a program accepting students with no computer science experience, but attracting many with significant experience from high school---and talent. 
Our main design goals/proposed solutions for the course were:
\begin{enumerate}
    \item address observed failure modes for design projects by having teams design games for older adults with a plausible mechanism for diagnosing Parkinson's Disease (data collection and analytics being beyond the scope of the course);
    \item accommodate students with varying backgrounds by giving teams a choice to develop single- or multiplayer games;
    \item balance the need for structure with the growth opportunity provided by flexibility via a set of fillable worksheets embedded in a PowerPoint editable/commentable by a student team and instructional staff. 
\end{enumerate}


\subsection{Human-Centred Design}
The HCD methodology is described in ISO 9241-210 as an ``approach to systems design and
development that aims to make interactive systems more usable
by focusing on the use of the system and applying human factors/ergonomics and usability 
knowledge and techniques''. This helps to create usable systems that
increase user productivity, reduce errors, reduce training and support needs,
and improve user acceptance~\cite{maguire2001methods}. 

Key to the application of human-centred design is using multidisciplinary
skills and perspectives, user-centred, evaluation-driven refinement of the
design, and involving the user throughout a highly iterative design 
process~\cite{giacomin2014human}.

\subsection{Design Thinking}
DT is an HCD approach popularized by
Tim Brown of IDEO in 2008~\cite{brown2008design}. 
Among Brown's messages is that design is not an innate ability,
but is rather ``the result of hard work augmented by a human-centred 
discovery process and followed by iterative cycles of prototyping, 
testing, and refinement''~\cite{brown2008design}. DT has been widely adopted
in industry, e.g. IBM's Enterprise Design Thinking~\cite{ibmdesignthinking}. We 
use the British Design Council's Double Diamond~\cite{designcouncil2021} to model 
the design process.

\subsection{Failure Modes}
We have observed four failure modes for student designers,
three of which have been observed by other authors:

\begin{enumerate}[label=FM\arabic*.,left=0.4cm]
    \item[FMSub] \label{FM:Sub}
    \textbf{Substitution}: Designers substitute their own feelings,
    wants or needs in place of the users'.
    For instance, it may be hard for youth to imagine older users' points of view. Kimbell~\cite{kimbell2011rethinking} remarks they often ``rely on hunches and presuppositions, not just facts''.

    \item[FMHN] \label{FM:HammerNail}
    \textbf{Hammer-Nail Problem}: While it may be necessary to restrict
    technology choices, including for course evaluation purposes, designers sometimes start with a particular technology and work backwards to the problem. Per Kimbell, this may be a failing of DT itself as DT ``continue[s] to privilege the designer, however empathetic, as the main agent in design''~\cite{kimbell2011rethinking}.
    
    \item[FMSC] \label{FM:SelfCensoring}
    \textbf{Idea Self-Censoring}: Designers may not contribute ideas if they
     feel they are inexperienced or less qualified than other team members.
    This is especially true for students from underrepresented groups, for instance female-identifying students, who are likely to have lower levels of
    self-efficacy and confidence~\cite{kijima2021using}.
    
    \item[FMJ] \label{FM:SimilarIdeas}
    \textbf{Jump to the First...}: There are several ways in which design teams may jump on their first viable problem definition or solution and fail to explore the design space.
    Elaborating on and being inspired by others' ideas is good,
    but if started too soon, other ideas may be lost or never discovered. 
    Similarly, lack of response during feedback sessions is not an excuse to stop iterating,
    it is more likely a sign that a prototype is failing even to draw criticism.  
\end{enumerate}

Three categories of scaffolding are needed to mitigate these failure modes: problem 
design, tool design, and course design. A combination of these three strategies is often necessary.

Problem design is of particular interest, and we present two example problems that we have
studied: 1) the Math Visualizer project and 2) the Parkinson's Disease (PD) detection game project. 
Throughout this paper, we will evaluate why we believe the latter to be a more effective problem.

\subsection{Paper Sections}
\textbf{\ref{Sec:Background}} discusses prior research 
on good problems for teaching, PD, and Event-Driven Programming (EDP). 
\textbf{\ref{Sec:OurApproach}} 
describes our approach to teaching design in a large first-year classroom, 
including the fillable worksheets and the PD problem.
\textbf{\ref{Sec:TEASync}} describes TEASync, our \emph{novel} multi-user EDP framework
which leverages Model-Driven Development (MDD) to lower the barriers for novice developers.  
\textbf{\ref{Sec:Methodology}} describes the context for the experiment and how it was evaluated. 
\textbf{\ref{Sec:Results}} presents results of surveys and focus groups and example projects. \textbf{\ref{Sec:ThreatsToValidity}} 
lists threats to validity. \textbf{\ref{Sec:Discussion}} then discusses and 
interprets the results. Finally, we discuss the generalizability of 
this approach and future work in \textbf{\ref{Sec:Conclusions}}.

\section{Background and Related Work}
\label{Sec:Background}

This section outlines work in designing good problems to
teach design as well as research motivating the Parkinson's
Disease (PD) game problem.

\subsection{Designing Good Design Thinking Problems}
\label{Sec:GoodProblem}

The quality of the problem is a critical factor in Problem-Based Learning (PBL) \cite{Lima2008EscreverBP, sockalingam2011characteristics}. Lima \cite{Lima2008EscreverBP} presents a model for creating effective problems, while Sockalingam \cite{Lima2008EscreverBP} outlines characteristics that contribute to desired learning outcomes. For insights into the traits of good problems, Reinholz and Valtanen offer valuable resources\cite{reinholz2018good,valtanen2012features}. Reinholz states that good problems share common attributes, such as real-world connections, reinforcing conceptual understanding, offering multiple solution paths, and a wide range of difficulty levels, among others~\cite{reinholz2018good}. Valtanen supplements this by examining problem features in an information technology context, emphasizing the necessity of a deep understanding of what defines a problem. Valtanen suggests good problems should lead to intended learning outcomes, generate interest, follow a suitable format, stimulate critical reasoning, promote self-directed learning, build upon prior knowledge, encourage elaboration, and foster teamwork \cite{valtanen2012features}.

Design thinking is an approach that can be utilized across different fields, improving both problem-solving and innovation in various industries \cite{DORST2011521}. The methodology of design thinking has significant potential to enhance the user experience in information science research, thus improving the overall user journey related to information \cite{Nakano2018}. Moreover, design and design thinking play a critical role in fostering creativity and innovation in STEM education and can be cultivated through design activities across different disciplines and integrated STEM education \cite{Li2019}. Despite numerous publications utilizing design thinking, there is a lack of focus on formulating appropriate problems for the DT process.

Many of the properties of good problems in the literature are covered by any 
design thinking exercise (e.g.~multiple solution paths, a wide range of difficulty 
levels, etc.). Therefore, in Fig.~\ref{Fig:GoodProblemProperties}, we propose 
three additional criteria to evaluate DT problems, along with the failure mode(s) 
they help to mitigate.

\begin{figure*}[]
    \centering
    \small
    \caption{Summary of Good Problem Properties}
    \label{Fig:GoodProblemProperties}
    \rowcolors{2}{gray!15}{white}
    \begin{tabular}{p{0.58\textwidth}p{0.12\textwidth}ccc}
        \toprule
      \textbf{Good Problem Property} & \textbf{Failure Modes} & \textbf{Source} & \textbf{MV} & \textbf{PD} \\
        \midrule
        \textbf{Engage empathy:} should involve users different enough from the learners & \textbf{FMSub}, \textbf{FMJ} & Proposed & \xmark & $\checkmark$ \\
        \textbf{Novel domain:} should involve a domain likely to be outside the learners' expertise & \textbf{FMSub}, \textbf{FMJ}  & Proposed & \xmark & $\checkmark$ \\
        \textbf{Users:} learners should have easy access to the target users, or good proxies & \textbf{FMSub} & Proposed & Mixed & $\checkmark$ \\
        \textbf{Options:} problem offers multiple valid solution paths & \textbf{FMJ} & \cite{reinholz2018good} & $\checkmark$ & $\checkmark$ \\
        \textbf{Real-world:} learners recognize given problem as a real-world problem &  & \cite{reinholz2018good} & $\checkmark$ & $\checkmark$ \\
        \textbf{Difficulty levels:} problem should be approachable to students of different levels of experience and ability & \textbf{FMSC} & \cite{reinholz2018good} & $\checkmark$ & $\checkmark$ \\
        \textbf{Critical reasoning:} problem encourages learners to think critically & \textbf{FMSub}, \textbf{FMSC}, \textbf{FMHam} & \cite{valtanen2012features} & $\checkmark$ & $\checkmark$ \\
        \textbf{Foster teamwork:} problem should allow natural division of labour within teams &  & \cite{valtanen2012features} & $\checkmark$ & $\checkmark$ \\
        \bottomrule
    \end{tabular}
    \vspace{1mm}

    Legend: MV = Math Visualizer Project; PD = Parkinson's Disease Project
\end{figure*}

\subsection{DT Problem: Games for Parkinson's Disease Detection}
\label{Subsec:GameDesignForOlderAdults}
According to the United Nations Population Prospect, by 2050, 1 in 6 people in the world will be over the age of 65, up from 1 in 11 in 2019 \cite{un2019w}.
PD is the second most prevalent neurodegenerative disorder globally, affecting about two percent of people over 65 years of age, four percent over 80 \cite{de2006epidemiology}, and more than one million people in North America \cite{lang1998parkinson}. The causes of the disorder remain elusive, with diagnostic methodologies that are currently incapable of definitively identifying PD, while therapeutic options for its management remain inadequate.
Therefore, early detection and monitoring are very important.
While games are commonly thought to be more attractive and enjoyable for younger people, studies have demonstrated that older individuals also engage in and derive pleasure from games \cite{de2011never,hall2012health}. They express interest in playing games when they perceive it to be a cost-effective way to support their health and enhance their overall well-being \cite{brown2012let, gerling2012game, lim2012closed}.

Many studies underscore the potential of smartphone technology in improving the management and understanding of Parkinson's disease. For example, Fraiwan et al. developed a system for mobile applications that can accurately detect hand tremors in Parkinson's patients \cite{fraiwan2016parkinson}. Pieter et al. explore the feasibility and effectiveness of smartphone-delivered automated feedback training for gait in Parkinson's patients, further demonstrating the potential of mobile technology in this field \cite{ginis2016feasibility}. Brian et al. and Arora et al. take this a step further by highlighting the potential of smartphones in detecting and monitoring symptoms, with Brian's mPower study using an iPhone app interface \cite{bot2016mpower} and Arora's pilot study using an Android app \cite{arora2015detecting}. Even though these works provide a solution for crowdsourcing data collection, they struggled to maintain volunteer engagement.
Perhaps because they did not emphasize co-design, in which their users participate in the development.

So the coming Silver Tsunami is more than a government problem, it is the perfect setup for a socially relevant problem with limitless possible solutions.
A real need exists for PD early diagnosis and treatment monitoring, 
and apps can play a role, but work needs to be done on engagement.
This creates an ideal problem area for budding designers,
evidently
ticking off \emph{engage empathy}, \emph{novel domain}, and \emph{real-world} in Fig.~\ref{Fig:GoodProblemProperties}.

\subsection{Event-Driven Programming and Model-View-Update}
EDP is a type of programming where concurrency is based around
the processing of events~\cite{dabek2002event}. This is instead of the use of IO in 
threads with blocking which
Dabek et al. \cite{dabek2002event} say introduces concurrent 
execution and locks where they are not needed.
In contrast, several authors show that the EDP paradigm leads to more reliable
software~\cite{dabek2002event, desai2013p}.
Despite EDP's benefits, new programmers still face challenges
while learning EDP, 
especially in reasoning about their program's 
behaviour,
according to Lukkarinen et al.~\cite{lukkarinen2021event} who recently surveyed the literature on EDP in education. 
While EDP in education research has been active since the 1990s,
they also concluded that (1) little research tries to separate the impact of EDP concepts from the 
GUI framework implementing it, which is usually the focus for teachers and students;
(2) insufficient research evaluates the impact of programming language,
in particular of functional programming languages; 
and (3) empirical research methods are needed to back up numerous experience reports.

\begin{figure}
    \centering
    \includegraphics[width=0.99\linewidth]{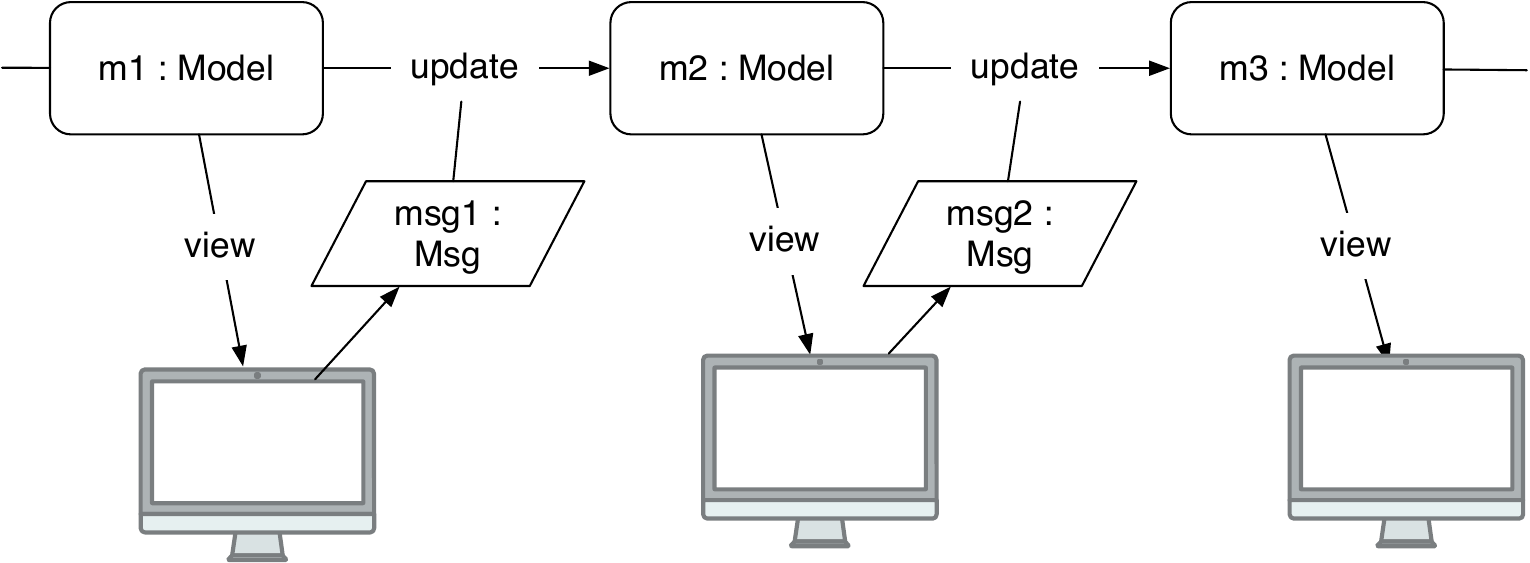}
    \caption{Model-View-Update dataflow.}
    \label{fig:MVU}
\end{figure}

Elm~\cite{czaplicki2012elm} is a functional language for web programming whose standard library is a clean implementation EDP.
The Elm Architecture (TEA)~\cite{farewellToFrp} was introduced to simplify
reasoning about EDP by defining two pure functions:
\begin{align}
\text{\elmi{update}}:\text{\elmi{Message}}\rightarrow\text{\elmi{Model}}&\rightarrow\text{\elmi{Model}},\ \text{and} \\
\text{\elmi{view}}:\text{\elmi{Model}}&\rightarrow\text{\elmi{Html}},
\end{align}
which apply events (messages) to calculate new states, and render the display from that state.
Calling the \emph{state} ``model'' fits TEA into the nomenclature for 
user-interaction paradigms starting with Model-View-Controller.
TEA is the exemplar of Model-View-Update (MVU).
Both \elmi{update} and \elmi{view} are pure \emph{functions}, not objects,
so user interaction can be understood as a data flow (see Fig.~\ref{fig:MVU}).
The lack of mutation makes it easier for beginners to reason about programs,
addressing perhaps the main concern raised by Lukarinen et al.
Prior offerings of our design course used Elm+TEA for this reason.
In addition to the innate advantages of a pure data flow for reasoning,
we adopted the graphical Model-Driven-Development (MDD) tool SD~Draw \cite{pasupathi2022teaching}.

\section{Large Classroom Design Thinking Teaching}
\label{Sec:OurApproach}

This section describes our approach to teaching DT in a large classroom
setting of approximately 200 first-year CS students. We will first discuss and evaluate
two types of DT problems we have tried, then explain our fillable worksheets
and our new Game Matrix method for brainstorming.

\subsection{This \emph{IS} Your Grandfather's Gaming App!}
Spurred by challenges with other DT problems wherein teams could choose target users too similar in life experiences to really exercise empathy, we piloted the Parkinson's Disease PD Detection Problem as a virtual camp, 
and made it the assigned problem for the design course in 2024.
Most undergraduate students have ready access to potential users in the form of grandparents and other older relatives 
who are invested in their success.
In a heterogeneous classroom, 
creating a video game for older adults provides a variety of development tasks from graphic design to game-AI design,
providing different difficulty levels and a large enough number of tasks to require teamwork.
Furthermore, teams are unlikely to
have much knowledge of Parkinson's Disease prior to the course, meaning
this was a \emph{novel domain} for them to reason about.
This leaves \emph{options} and \emph{critical reasoning} to be developed by the design method
as prescribed in the fillable worksheets.
Furthermore, working for older adult clients makes it difficult for young people to 
substitute their own experience for the user's, avoiding \textbf{FMSub}.

\subsection{Fillable Worksheets, featuring the Game Matrix}
Although DT is an iterative, adaptive process, for the purpose of a
course we have found it necessary and helpful to ``unroll the loop'',
to walk students through a common set of iterations,
introducing concepts along the way,
and reinforcing design thinking using examples from team updates. 
We give students an overview of the process and set expectations
by sharing a fillable
worksheet template, delivered as a Microsoft PowerPoint slide deck. 
Table~\ref{Fig:SlideTemplate} describes each slide in the
slide worksheets given to the groups.
Many slides are duplicated to set expectations for the number of iterations required.
Many students were initially surprised by the number of iterations expected.
In a real, open-ended design project of this type, they would be doing many more.
By the end of the course, we observed that
poorly performing teams were scrambling to meet check-in deadlines and failed to plan prototypes as experiments to deepen their understanding of their user and their needs. 
Some worksheets, like the Empathy Map, are common to almost all implementations of DT.
A few, notably the Game Matrix, 
are novel,
and others, like the  
Desirability vs. Feasibility plot 
(Fig.~\ref{Fig:FeasbilityDesirability}) 
are adaptations of common tasks to worksheet form.

\medskip
Brainstorming has been shown to be an effective tool to help promote creativity
in design, but different strategies can affect what parts of the design process groups focus on~\cite{gero2013design}.

We developed the \emph{Game Matrix} to address an observed failure to explore the problem or solution space before settling on a problem/solution (\textbf{FMJ}),
by separating the ideation problem into 16 separate problems. 
The Game Matrix (Fig.~\ref{fig:GameMatrix}) consists of two axes, which should be non-overlapping views of the problem,
and are preferably lined up with research the learners do into the 
problem or based on ideas generated by the initial interviews. For
 the PD problem, the columns represent symptoms researched by the teams,
 and rows represent user interests gleaned from interviews.
The number of rows and columns can be adjusted to the problem,
but we have found that in this format, teams will rise to the challenge of
articulating 16 concepts for games which could be used to detect PD.
The matrix structure forces the ideas to span a bigger solution space than we have observed student teams exploring previously, due to \textbf{FMHN}, \textbf{FMSC}, and \textbf{FMJ}.

\begin{figure}[h!]
\centering
\caption{The Game Matrix}
\label{fig:GameMatrix}
\begin{tabular}{p{0.85cm}|p{1.25cm}|p{1.25cm}|p{1.25cm}|p{1.25cm}|}
\cline{2-5}
                                   &  {\small SymptomA} &  {\small SymptomB} & {\small SymptomC} & {\small SymptomD}\\ \hline
\multicolumn{1}{|l|}{Interest 1}   &            &            &        &    \\ \hline
\multicolumn{1}{|l|}{Interest 2}  &           &            &         &   \\ \hline
\multicolumn{1}{|l|}{Interest 3} &           &            &         &   \\ \hline
\multicolumn{1}{|l|}{Interest 4} &           &            &         &   \\ \hline
\end{tabular}
\end{figure}

\medskip
Students used the worksheets to plan their work, record their decisions, and track their progress. 
They were encouraged to use their worksheets as visual aids for project updates.
A total of nine
prototypes were required throughout the term, approximately one
or two per week. Four DT updates were due throughout the term. DT updates
are informal, 5-minute videos, submitted approximately every two weeks,
where students share their worksheets and describe what they have worked
on or modified since the last update as a voiceover.
They were evaluated based on content and not production 
value (e.g. the majority submit an unedited Microsoft Teams video call recording).
The combination of recorded video updates, written responses and lectures devoted to common feedback on the updates
was the most effective way to reinforce \emph{critical thinking},
as observed by interactions with students in office hours and labs.
Design \emph{options} were called for in many places,
but we have found it necessary to \emph{require} two independent paper prototypes
in the first stage of paper prototyping,
to force design teams to get feedback on multiple ideas and discuss the design space,
even after they have chosen a solution, again to avoid \textbf{FMJ}.
After completing the first two prototypes, we ask them to evaluate whether a pivot is in order,
in which case they would duplicate worksheet slides going back to at least the initial prototyping, but possibly as far back as initial interview question formulation.

\begin{figure}
    \centering
    \fbox{\includegraphics[width=\linewidth]{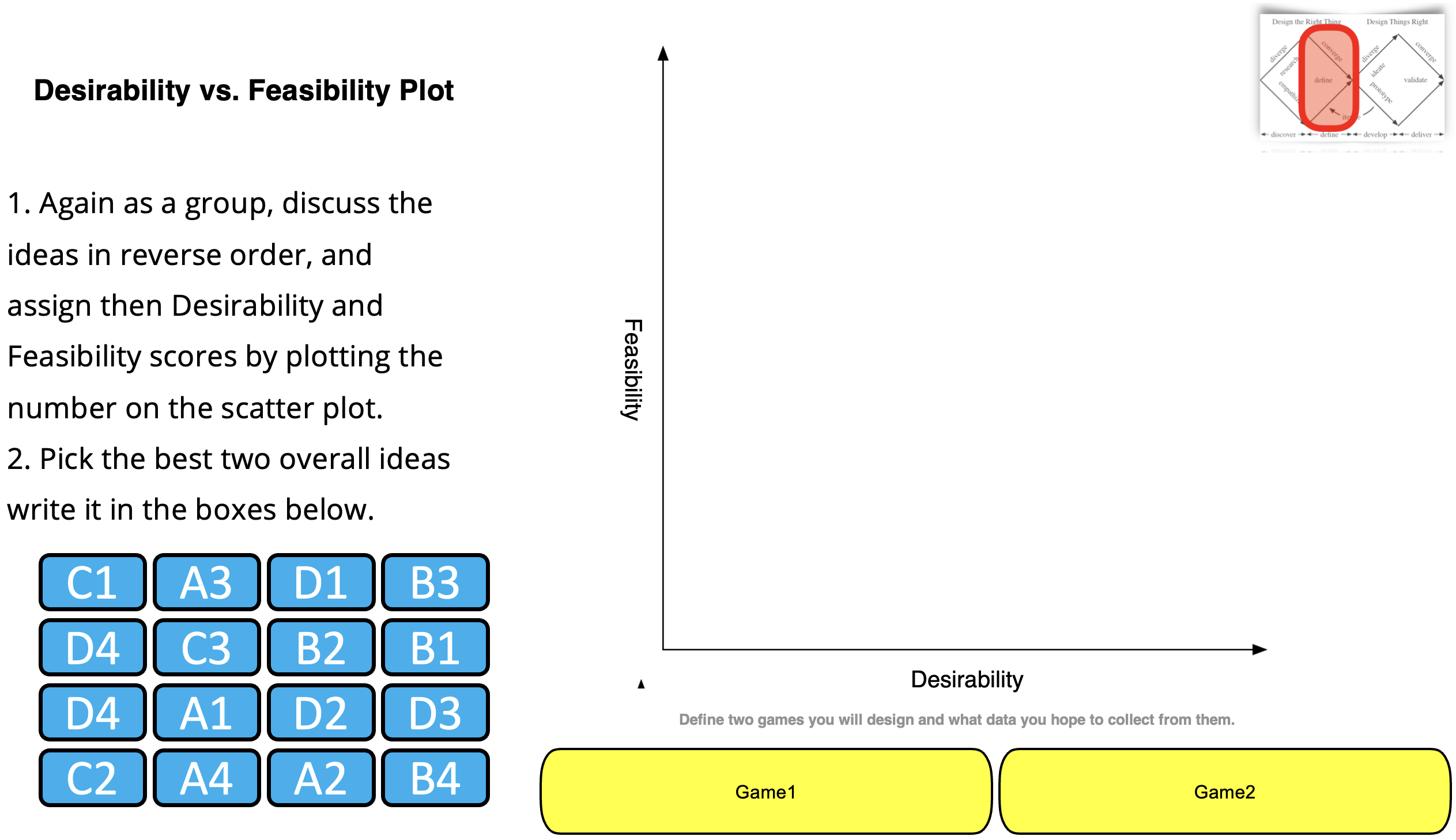}}
    \caption{Desirability vs. Feasibility worksheet. Students rate their 
    ideas from the Game Matrix (Fig.~\ref{fig:GameMatrix}) and drag them onto the plot to choose their top 2 ideas to pursue.
    The plot has some value as a product,
    but its main value is in engaging \emph{teamwork} by forcing the design team to discuss the placement of different solutions, 
    and in \emph{critical thinking}.
    }
    \label{Fig:FeasbilityDesirability}
\end{figure}

\begin{table*}[]
    \small
    \centering
    \caption{Overview of Fillable Design Thinking Worksheets}
    \label{Fig:SlideTemplate}
    \rowcolors{2}{gray!15}{white}
    \begin{tabular}{p{0.25\textwidth}p{0.70\textwidth}}
        \toprule
      \textbf{Slide} & \textbf{Description}\\
        \midrule
      Title Slide & Includes group number, name, and student names.\\
      Double Diamond & Double Diamond graphic showing overview of the DT process. \\
      Timetable & A schedule of deliverables, showing where learners should be in the slides throughout the course.\\
      Understanding Our User 1 \& 2 & A few research questions to 
      prime the learners about two possible target users.\\
      Sensitivity \& Privacy Reminder & An optional reminder to be sensitive and protect privacy in 
      interviews, which is especially necessary for problems about 
      sensitive topics like the Parkinson's problem.\\
      Introduction Script & A slide for groups to write their pre-interview script to be read, explaining the purpose of their project and how the interview will take place.\\
      Questions to Ask & A slide for groups to create questions to ask in their interviews.\\
      Interview Transcript & The group's raw notes from their interview.\\
      Empathy Map (EM) & A 4-square grid containing ``say'', ``think'', ``do'', and ``feel'', where groups categorize the raw notes from their interview. \\
      Interview Skills Evaluation & A rubric for groups to perform a self-evaluation of their interview skills.
      \\
      Revised Questions & After performing their mock interview, a list
      of revised questions based on problems they found.
      \\
      Transcript / EM / Evaluation (x4) & Four more copies of the aforementioned slides are provided to allow for more interviews.
      \\
      Research Slides & Space for groups to research the problem, e.g.
      Parkinson's Disease symptoms. These will be the basis for one axis of the 
      Game Matrix.
      \\
      Users' Interest & Space for groups to write down all interests that
      the user may have stated in the interviews. Can be used as the other axis for the Game Matrix.
      \\
      Game Matrix & A grid for structured brainstorming, with two disjoint axes. For projects not involving games this could be generalized and called the ``Brainstorming Matrix''. See Fig.~\ref{fig:GameMatrix}.
      \\
      Desirability vs. Feasibility Plot & A plot of all ideas in the Game Matrix according to the group's perception of the user's desirability of the solution and the feasibility of implementing the solution. The team picks their two best ideas according to this plot to start creating prototypes. Example shown in Fig.~\ref{Fig:FeasbilityDesirability}.
      \\
          Discussion Points & Thoughts the group may have that may be useful later, e.g. tech or graphics they may need.
      \\
      Prototype 1 & Overview of (paper) prototype 1. More slides can be added if necessary. 
      \\
       Prototype 2 & Overview of (paper) prototype 2. More slides can be added if necessary. 
      \\
      Prototype 1 Empathy Map (Mock) & An empathy map for a mock interview about prototype 1, e.g. with classmates.
      \\
      Action Plan (Mock) & A table of improvements needed in successive prototypes.
      \\
      Prototype 1 Empathy Map & An empathy map from their prototype 1 demonstration interview.
      \\
      Prototype 2 Empathy Map & An empathy map from their prototype 1 demonstration interview.
      \\
      Prototype 1 \& 2 Pros and Cons & A table of pros and cons, to select a prototype to continue with.
      \\
      Action Plan & A table of improvements needed in successive prototypes.
      \\
       Prototype 3 & Overview of (paper) prototype 3. More slides can be added if necessary. 
      \\
       Technical Challenges & Having settled on one idea, the group can start to address anticipated technical challenges. 
      \\
       Peer Feedback & Groups should present their work thus far to peers and take notes.
      \\
       Prototype 3 Feedback Grid  & Feedback grid from the prototype 3 demonstration interview.
      \\
       Prototype 3 Action Plan & Action Plan from the prototype 3 demonstration interview.
      \\
       Prototype 4 & Overview of (paper) prototype 4. More slides can be added if necessary. 
      \\
       Risks & Slide for groups to reflect on risks that may preclude a successful project.
      \\
      
       Prototype 4 Feedback Grid  & Feedback grid from the prototype 4 demonstration interview.
      \\
       Prototype 4 Action Plan & Action Plan from the prototype 4 demonstration interview.
      \\
       Prototype 5-9, with Feedback Grids and Action Plans (many slides) & Slides for groups to create prototypes 5-9, and fill out empathy maps and Action Plans. By prototype 5 or 6, groups are expected to have started implementing functional prototypes of concepts/features. Integration starts in prototype 7 or 8.
      \\
       Project Pitch Slides & Slides for groups to use as their final pitch. These include introducing the user, their problem, why they are qualified to help the user, the solution, and what the team learned from the DT process.
       Instructions emphasize the need to communicate the \emph{critical thinking} they engaged in.
      \\
        \bottomrule
    \end{tabular}
\end{table*}

\section{TEASync:  A Framework for Multi-User Games}
\label{Sec:TEASync}

\begin{figure}
    \centering
    \includegraphics[width=\linewidth, trim=25cm 0 1.5cm 0, clip]{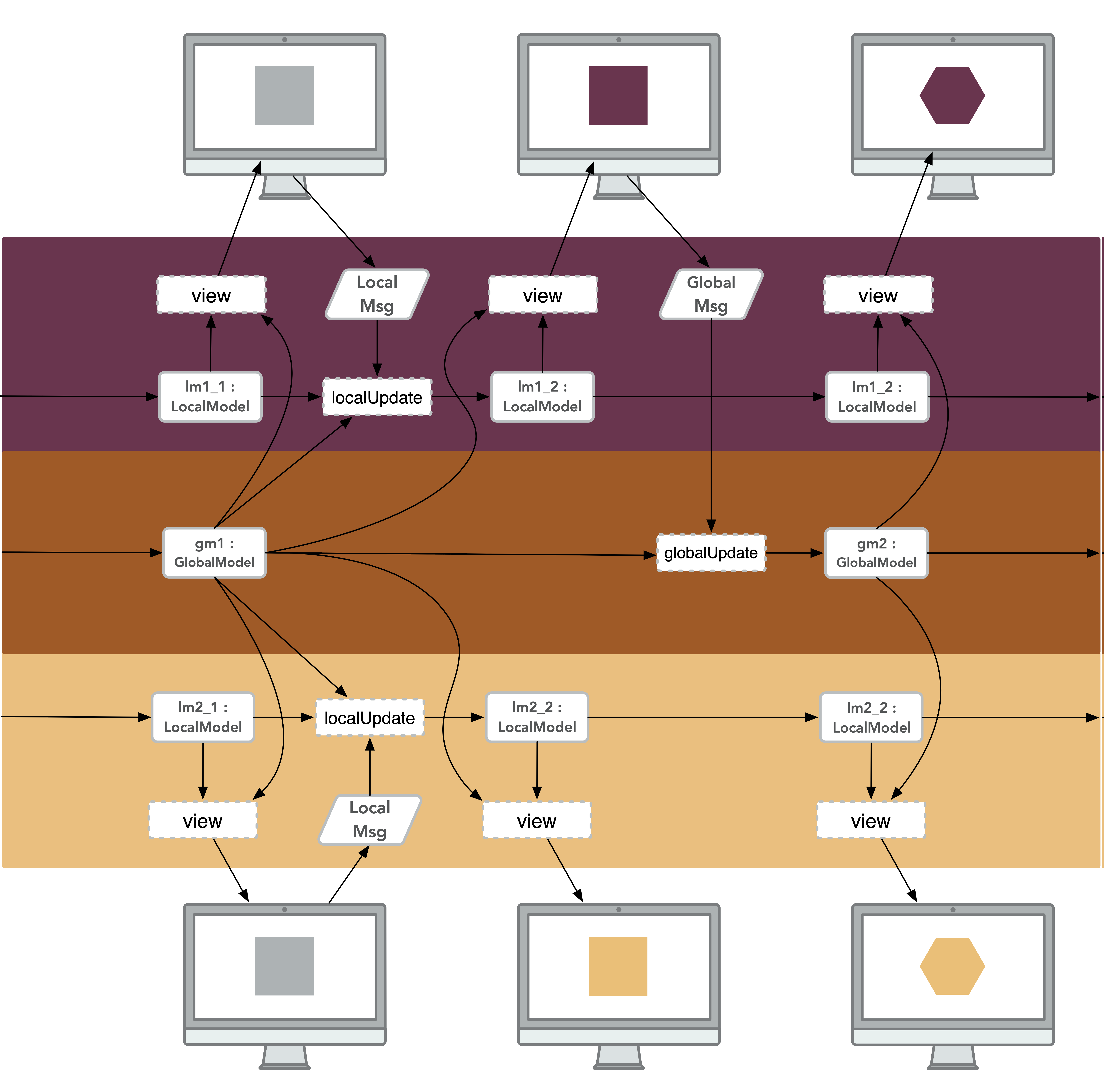}
    \caption{Local-Global Model-View-Update (LG-MVU) architecture dataflow diagram.
    This simple example stores a colour in the local model and the shape itself in the 
    global model. Thus the colour can be changed by local messages and shape change 
    global messages are propagated to all clients.}
    \label{Fig:LG-MVU}
\end{figure}

To address the huge variation in experience,
we developed a new EDP framework inspired by TEA, which we call TEASync.
Unlike TEA which is an MVU framework,
we invented a new classification \textit{Local-Global Model-View-Update} (LG-MVU) to describe it.
Just as MVU makes it easy to reason about single-user applications as a pure data flow (without side effects),
LG-MVU brings the same transparency to distributed, multi-user applications.
Given our audience, it was important to express such client-server systems in Elm alone,
without a separate language for  server-side code.
In fact, the role of the server is minimized by presenting the entire application
as a pure data flow with layers.
Fig.~\ref{Fig:LG-MVU} shows a dataflow diagram for an LG-MVU app with two clients. In
this simple app example, the two clients see the same shape, but each client can set its own colour.
The display is rendered in the figure inside the computer displays.
Considering only the top (purple) or bottom (gold) layers,
the data flow reduces to the MVU data flow in Fig.~\ref{fig:MVU}.
The central layer adds a global model which holds state globally accessible/affecting all clients.
Unlike the local layers (of which there could be any number) the global thread has no attached view,
however, every local user can generate messages (events) processed by the global thread.

Obviously, it is necessary to synchronize this global state,
and to order the global events, so every client maintains a consistent state.
To handle this the TEASync framework includes Integrated-Development-Environment (IDE) support
in which the IDE parses global data types (and child types) and generates type-safe encoders and decoders
so that the TEASync runtime can maintain a single global state duplicated, and synchronized across clients.

Technical details of the TEASync framework are currently being written up as a Master's
Thesis.

\section{Methodology}
\label{Sec:Methodology}

The current study focuses on a first-year Computer Science ``Intro to Software Design'',
attended by approximately 200 students. 
The course's focus was the design of a game appealing to older adults which could plausibly be used for PD diagnosis. 
The emphasis was on being appealing to older adults.
Self-selected teams of 4-7 students spent the first month learning EDP using the MDD approach enabled by SD~Draw, and in parallel developed interview questions, interviewed older adults, developed empathy maps, ideated and ranked solutions and developed two paper (i.e., non-functional) prototypes drawn in PowerPoint slides.
Their first midterm included a screenshot of a state diagram for a simple adventure game and the code generated by SD~Draw, with specifications for a point system requiring many coordinate edits.
While adventure games are very relatable, and recommended as a teaching example,
due to high levels of mastery demonstrated on this midterm,
this choice probably helped steer most groups to develop an overall adventure game with multiple mini-games. 
After the first month, 
TEASync was introduced and a third midterm required augmenting a distributed cash-register/inventory application.
Mastery was not reached by most students,
which was disappointing but okay given the goal of providing a range of challenges to match the heterogenous cohort.
At this point, teams faced the design decision: whether to develop a single- or multiplayer app.
Given the semester length, and desire to start the design process early,
the knowledge they needed to make this decision forced the decision later in the design process 
when many groups had already made substantial progress on designing a single-user game.

The course was allocated 2.5 hours of  
teaching assistant (TA) contact hours per student, 
plus 0.5 hours/student of marking support. 
Groups had four hours of lab per week to get help from TAs and ask questions,
plus two hours of lecture and an office hour with the instructor. 
For consistency in marking and feedback, the course instructor marked DT updates and final deliverables. 

\subsection{Evaluation Methods}
Students were invited to complete two surveys, a focus group, and code 
compilation data analysis. Students
could opt into or out of any of these instruments without penalty. 
For completing the surveys and data collection, students had the option
to be entered into a draw to win a \$20 gift card of their choice. 
Students participating in the one-hour focus group were given a free 
pizza lunch upon completion.

The pre-implementation survey was given to students approximately 4 weeks
into the course, prior to starting to implement the project, but after
they had started the DT process. The post-implementation survey was given
in the second-last week. The pre-implementation survey 
consisted of 19 questions (including 6 for informed consent), and
the post-implementation survey consisted of 14 questions (including 7 
for informed consent). For Likert scale questions, options 
were assigned numeric values to calculate means.

The focus groups (two one-hour sessions with the same script) took place in the 
third-last week of the term, during labs. For the focus group, students were not 
explicitly identifiable. Transcripts were automatically generated by
the Microsoft Teams platform and were corrected by hand from the 
recording as needed. The study was approved by the McMaster Research Ethics Board,
as project \#6868. Unless otherwise indicated, quotations were collected in one 
of the two focus groups.

\section{Results \& Student Feedback}
\label{Sec:Results}

We collected feedback from students via surveys and focus groups. Occurrences of 
filler words such as ``umm'' and ``like'' have been removed for readability.

\medskip

It was encouraging that in spite of the challenges of working in groups and the increase in workload at university relative to high school,
students felt the effort was worthwhile:
\begin{quote}
``That's kind of cool to see something you build come to life and see people actually enjoy the products you make.''    
\end{quote}
and that, at least for most groups, peer collaboration made things better:
\begin{quote}
``[...] collaboration aspect of working all together and filming our DT Update was pretty fun [...]''
\end{quote}
One student even said that the best part of the course was
\begin{quote}
``just working with your groupmates.''
\end{quote}

\subsection{Teamwork}

Some students discussed the challenges of organizing a team:
\begin{quote}
    ``It was really hard to get everyone to do something and parallelize all our work. [...] So the biggest challenge was definitely organizing that.''
\end{quote}
which mirrors the experience of the instructional team,
who did need to help many groups with small and some large organizational issues.

\subsection{Empathize}

We chose the PD screening problem so students would recognize the need for interviewing,
because their own experience would be very different from their target users.  
Focus group participants not only discovered this aspect of the problem, but they came to enjoy it: 
\begin{quote}
    ``I specifically really enjoyed the aspect where we had to design games for demographic completely outside of our own because it really sort of forced us to engage with the differences and the different playing styles that different people have and also really engage the variety of even within a demographic.''
\end{quote}

Several students talked about not being able to take shortcuts, or fake the interview process:
\begin{quote}
    ``I think it's a helpful introduction to project design and also to the importance of thinking because if it was just create [a product] for our demographic, then we could just use a lot of our own ideas.''
\end{quote}

Although all students described interviewing users as a new task for them, 
some realized, in hindsight, that it was a useful skill to have:
\begin{quote}
    ``[I]f I work for a company, there's obviously gonna be a client.
And so that's gonna be very useful to be able to handle interviewing[.]''
\end{quote}

Although our choice of problem forced them to empathize,
with many alluding to having to think hard about how to explain concepts unknown to 
their audience,
interviewing their own grandparents, relatives, or neighbours
gave them a largely supportive user group for their first interviews.
It's not surprising that many students made comments similar to
``I really enjoy that aspect.''

While the difference in points of view was probably most obvious at the initial interview,
we know that they were actively engaging with their users by the comments they made 
related to feedback sessions,
For example, the difficulty of
\begin{quote}
``trying to explain the concept of a video game to somebody who's never really played a video game before.''    
\end{quote}
Similarly, another group reported
\begin{quote}
``what we think is intuitive wouldn't be as intuitive to them, and trying to figure out how to explain it in terms that they understand was a little bit challenging''    
\end{quote}

They were also prepared to be surprised and change their designs. One group reported:
\begin{quote}
``We found that the interviewees really liked the social aspect, and we thought it would be a good idea for older people, especially seeing [...] what happened during COVID [...]''    
\end{quote}
Other groups found the opposite, that social interaction was important, but not something they looked for in a video game.
Again and again, design teams used these reactions to reprioritize features,
which is exactly as we hoped.

\subsection{Game Matrix}

When asked to compare this ideation process with previous experiences, 
some students were very aware of \textbf{FMJ}:
\begin{quote}
``I think with regular brainstorming, what I found in the past was that once you kind of get an idea that [...] you kind of stop and then you just start with that right away.  But with [the Game Matrix] we were forced to finish the chart and then we ended up finding more good ideas later on when we were creating the chart.''
\end{quote}
This fits with the observations of the instructors that discussion was wide-ranging and involved the whole group, during this part of the lab. 

Another group did not end up using an idea from the Game Matrix, but reported that ``it forced us to be creative'', which started a process where
``everybody's building on top of everybody else's ideas.''
This is consistent with our theory that restricting ideas and requiring ideas in multiple categories forces new designers to stop censoring their ideas, avoiding \textbf{FMSC} 
and gets the whole team contributing. 

Another student went further and expects to continue using this approach because 
\begin{quote}
``[...] even beyond this course, [the Game Matrix] method does have some useful applications, I remember coming up with an idea or two that I happened to actually quite like that I might use as a future project.''    
\end{quote}
Asked to clarify whether it was just the ideas, or the process itself which could be reused,
they explained that they would 
\begin{quote}
``just lay down a grid of possible prompts and random words, and maybe a little mechanics, and then force yourself to think of a game idea built around that.''
\end{quote}

\subsection{Reaction to the Scaffolded Design Thinking Process}

Overall, students found value in the DT worksheets with many small steps:
\begin{quote}
    ``I actually really enjoyed the rigor of the whole process.
I like how it always gave us something to do, so we were always sort of thinking about the project and tapping into the project and just progressing with it.''
\end{quote}

On the flip side,
with a slide/worksheet devoted to every individual step,
\begin{quote}
    ``you have to search for things a lot and especially cause it's 100 or so slides it was difficult to find what you were looking for in all the PowerPoint slides.''
\end{quote}
This illustrates one disadvantage of using a general-purpose tool,
such as PowerPoint, for editing the worksheets.
Although search within general-purpose documents will likely keep improving, there is an opportunity for a purpose-built tool which makes searching for information easier.
But design teams would not have experienced this pain point if they 
were not following the process by 
 justifying their actions using references to things users said in interviews. 

Another danger of being too prescriptive, 
especially with busy undergraduates, is that they will not take advantage of flexibility.
One example would be a student reporting that
\begin{quote}
``it felt like there was a lot of planning before we actually tried out ideas and I feel like it might have been beneficial if we had more opportunities to actually I guess try things before earlier on cause to actually see if an idea is good.''
\end{quote}

Focus group participants were very engaged by this topic, and had several suggestions for improvements, 
including focusing on a different activity each week, 
adding better version control to the worksheets so that teams could see what was added during a particular interview/prototype cycle rather than having to look at track-changes for one slide at a time, and adding a reminder system to track which tasks were complete/overdue/up next.

With the perspective of having finished the project,
one student wrote in the post-survey that 
\begin{quote}
``Working through iterative prototypes and design thinking definitely made coding a big project feel complete rather than rushing through it for a due date. I also think developing a game is actually very interesting for a project as we can be creative and have real life applicability for it.''
\end{quote}

Finally, we explained the long-term advantage of getting feedback early and often, 
using paper prototypes as much as possible,
and forced them to practice this with early and numerous prototypes,
for which they needed to gather feedback and report on it in design updates.
But having a common set of deliverables meant that some groups were collecting data based on paper prototypes even when their concept was difficult to explain in a paper prototype:  
\begin{quote}
``we're trying to translate an idea we had in our heads that we think they might like to someone who can't directly interact with the program themselves.''    
\end{quote}

\subsection{Survey Results}
\label{SubSec:SurveyResults}

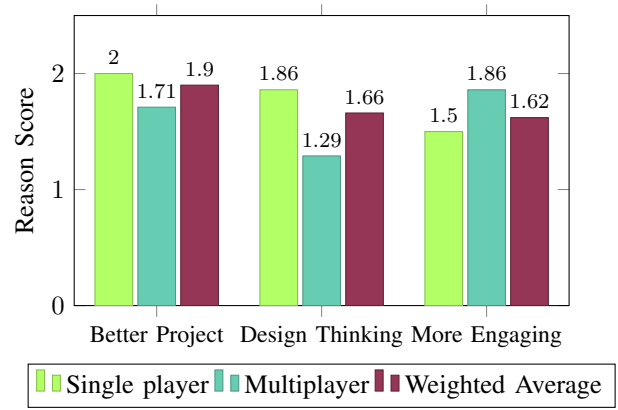
\begin{figure}
    \centering
    \begin{tikzpicture}
        \begin{axis}[
            ybar,
            bar width=.5cm,
            width=0.45\textwidth,
            height=.3\textwidth,
            enlarge x limits=0.25,
            symbolic x coords={Better Project, Design Thinking, More Engaging},
            xtick=data,
            x tick label style={anchor=north, font=\small},
            ymin=0, ymax=2.5,
            ylabel={Reason Score},
            legend style={at={(0.5,-0.2)}, anchor=north, legend columns=-1},
            nodes near coords,
            every node near coord/.append style={font=\footnotesize},
        ]
        \addplot[draw=ELMpos!75!black, fill=ELMpos] coordinates {
            (Better Project, 2.0)
            (Design Thinking, 1.86)
            (More Engaging, 1.5)
        };
        \addplot[draw=ANON!75!black, fill=ANON] coordinates {
            (Better Project, 1.71)
            (Design Thinking, 1.29)
            (More Engaging, 1.86)
        };
        \addplot[draw=ROUNDneg!60!black, fill=ROUNDneg] coordinates {
            (Better Project, 1.90)
            (Design Thinking, 1.66)
            (More Engaging, 1.62)
        };
        \legend{Single player, Multiplayer, Weighted Average}
        \end{axis}
    \end{tikzpicture}
    \caption{Top three reported reasons for choosing single vs. multiplayer [n=21]. Better Project = ``We thought we could create a better project by making a [single player/multiplayer] game'', Design Thinking = ``We thought a [single player/multiplayer] game would be more applicable to the Design Thinking problem'', More Engaging = ``We thought the resulting game would be more engaging being [single player/multiplayer].'' Reported here are the mean values from a Likert scale question of agreement with the statement.
    }\label{Fig:TopReasons}
\end{figure}

Fig.~\ref{Fig:TopReasons} gives the top three stated reasons for the students'
technology (single versus multiplayer) choice for their project. These 
three reasons were 
``We thought we could create a better project by making a [single 
player/multiplayer] game'', ``We thought a [single player/multiplayer] game would 
be more applicable to the Design Thinking problem'', and ``We thought the 
resulting game would be more engaging being [single player/multiplayer].'' 
Students answered on a Likert scale including four options: ``Does not apply to 
us'', ``Somewhat applies to us'', ``Applies to us'', and ``Completely applies to 
us''. These were assigned the values 0, 1, 2, 3 and then means were calculated.
Notably, these three reasons were
related to improving the project, rather than options related to preferences or 
technology such as ``We wanted to try something new''.

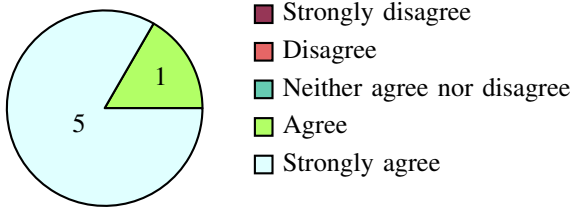
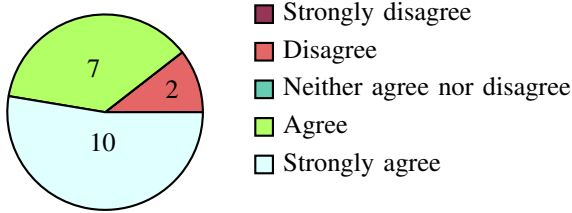
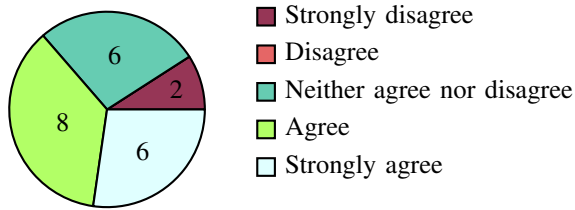
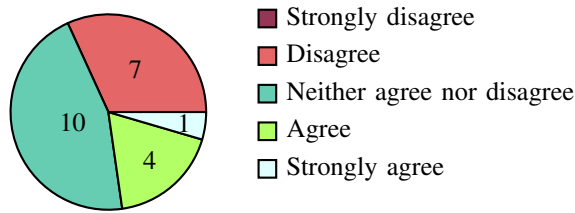
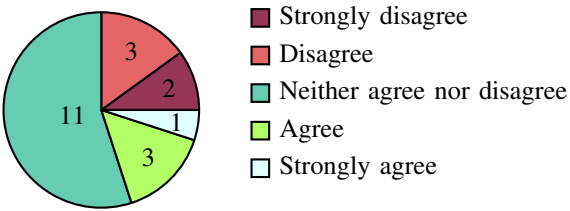
\begin{figure}
\begin{subfigure}{0.43\textwidth}
    \centering
    \begin{tikzpicture}
        \pie[
            text=legend,
            radius=1.3,
            color={ROUNDneg, ELMneg, ANON, ELMpos, CONCISE},
            sum=auto
        ]{
            0/Strongly disagree,
            0/Disagree,
            0/Neither agree nor disagree,
            1/Agree,
            5/Strongly agree
        }
    \end{tikzpicture}
    \caption{"It is important for software professionals to understand user needs." (Female-identifying respondents) [n=6]}
    \label{Fig:ImportantFemale}
\end{subfigure}

\begin{subfigure}{0.43\textwidth}
    \centering
    \begin{tikzpicture}
        \pie[
            text=legend,
            radius=1.3,
            color={ROUNDneg, ELMneg, ANON, ELMpos, CONCISE},
            sum=auto
        ]{
            0/Strongly disagree,
            2/Disagree,
            0/Neither agree nor disagree,
            7/Agree,
            10/Strongly agree
        }
    \end{tikzpicture}
    \caption{"It is important for software professionals to understand user needs." (Male-identifying respondents) [n=19]}
    \label{Fig:ImportantMale}
\end{subfigure}

\begin{subfigure}{0.43\textwidth}
    \centering
    \begin{tikzpicture}
        \pie[
            text=legend,
            radius=1.3,
            color={ROUNDneg, ELMneg, ANON, ELMpos, CONCISE},
            sum=auto
        ]{
            2/Strongly disagree,
            0/Disagree,
            6/Neither agree nor disagree,
            8/Agree,
            6/Strongly agree
        }
    \end{tikzpicture}
    \caption{"The TEASync framework is a useful tool for creating multiplayer games" [n=22]}
    \label{Fig:UsefulTool}
\end{subfigure}

\begin{subfigure}{0.43\textwidth}
    \centering
    \begin{tikzpicture}
        \pie[
            text=legend,
            radius=1.3,
            color={ROUNDneg, ELMneg, ANON, ELMpos, CONCISE},
            sum=auto
        ]{
            0/Strongly disagree,
            7/Disagree,
            10/Neither agree nor disagree,
            4/Agree,
            1/Strongly agree
        }
    \end{tikzpicture}
    \caption{"The TEASync framework was difficult to learn" [n=22]}
    \label{Fig:DifficultToLearn}
\end{subfigure}

\begin{subfigure}{0.43\textwidth}
    \centering
    \begin{tikzpicture}
        \pie[
            text=legend,
            radius=1.3,
            color={ROUNDneg, ELMneg, ANON, ELMpos, CONCISE},
            sum=auto
        ]{
            2/Strongly disagree,
            3/Disagree,
            11/Neither agree nor disagree,
            3/Agree,
            1/Strongly agree
        }
    \end{tikzpicture}
    \caption{"The TEASync framework was powerful enough for what we wanted to do" [n=20]}
    \label{Fig:PowerfulEnough}
\end{subfigure}
\caption{Likert scale questions results. (a) and (b) are from the pre-project survey,
and the rest are from the post-project survey.}
\end{figure}

Figs.~\ref{Fig:ImportantFemale} and~\ref{Fig:ImportantMale} show the
response to the prompt ``It is important for software professionals to
understand user needs'' from the pre-survey.
The agreement was stronger for female-identifying respondents, albeit
at a lower sample size.

Finally, Figs.~\ref{Fig:UsefulTool},~\ref{Fig:DifficultToLearn}, 
and~\ref{Fig:PowerfulEnough}, show results for the TEASync framework in terms of
students' ratings for TEASync's usefulness, difficulty to learn, and being powerful
enough to achieve what they wanted to.

\medskip

\begin{figure}
    \centering
    \includegraphics[width=\linewidth]{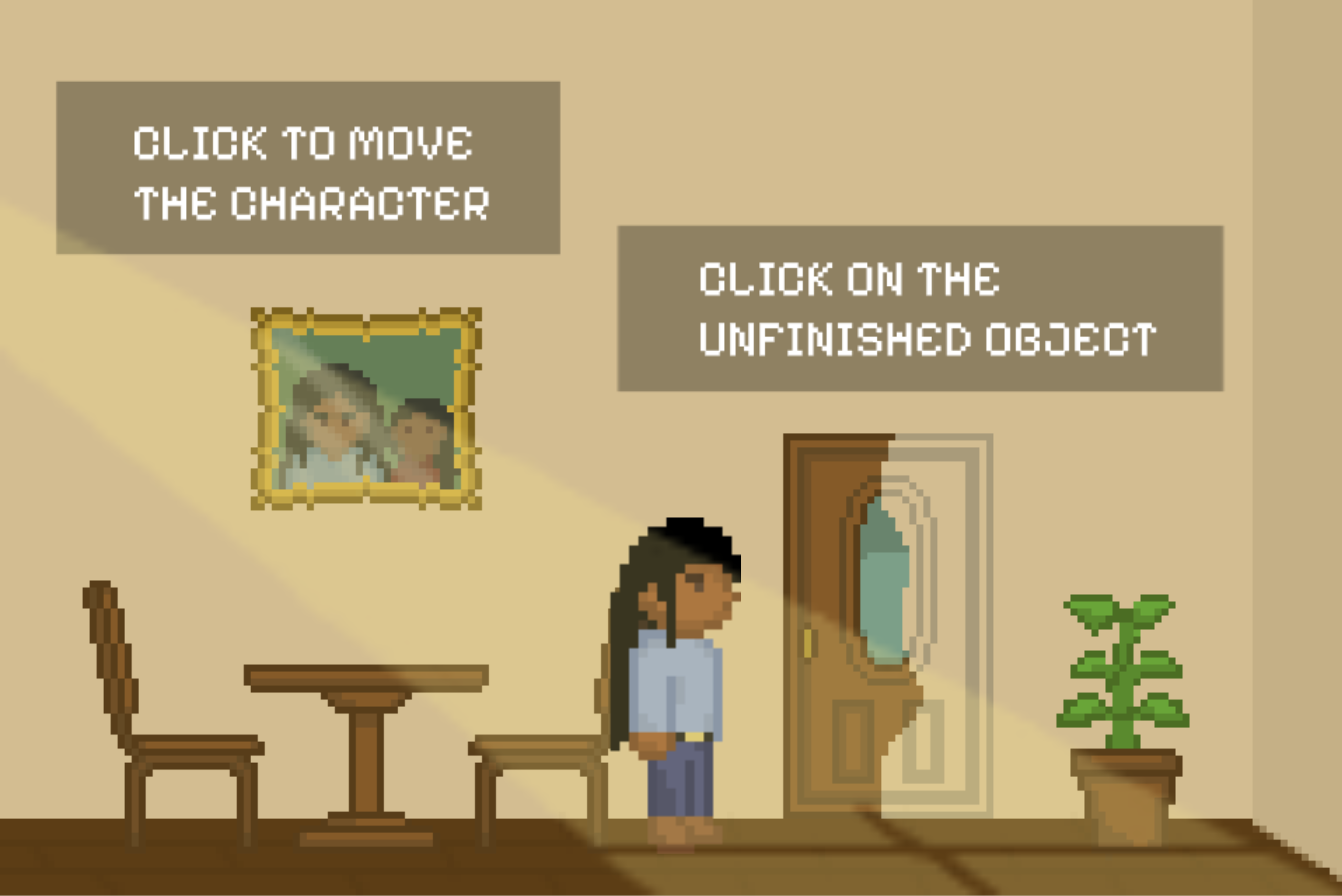}
    \caption{The single player game ``Brushstroke Journey'', created by first-year CS students. Players fill in objects by painting pixels, and mouse data can be collected to detect PD symptoms such as tremors. Although the game was 
    single-player, the group used the multiplayer framework to add features like a
    leaderboard (a common request from users).}
    \label{Fig:BrushstrokeJourney}
\end{figure}

\begin{figure}
    \centering
    \includegraphics[width=\linewidth]{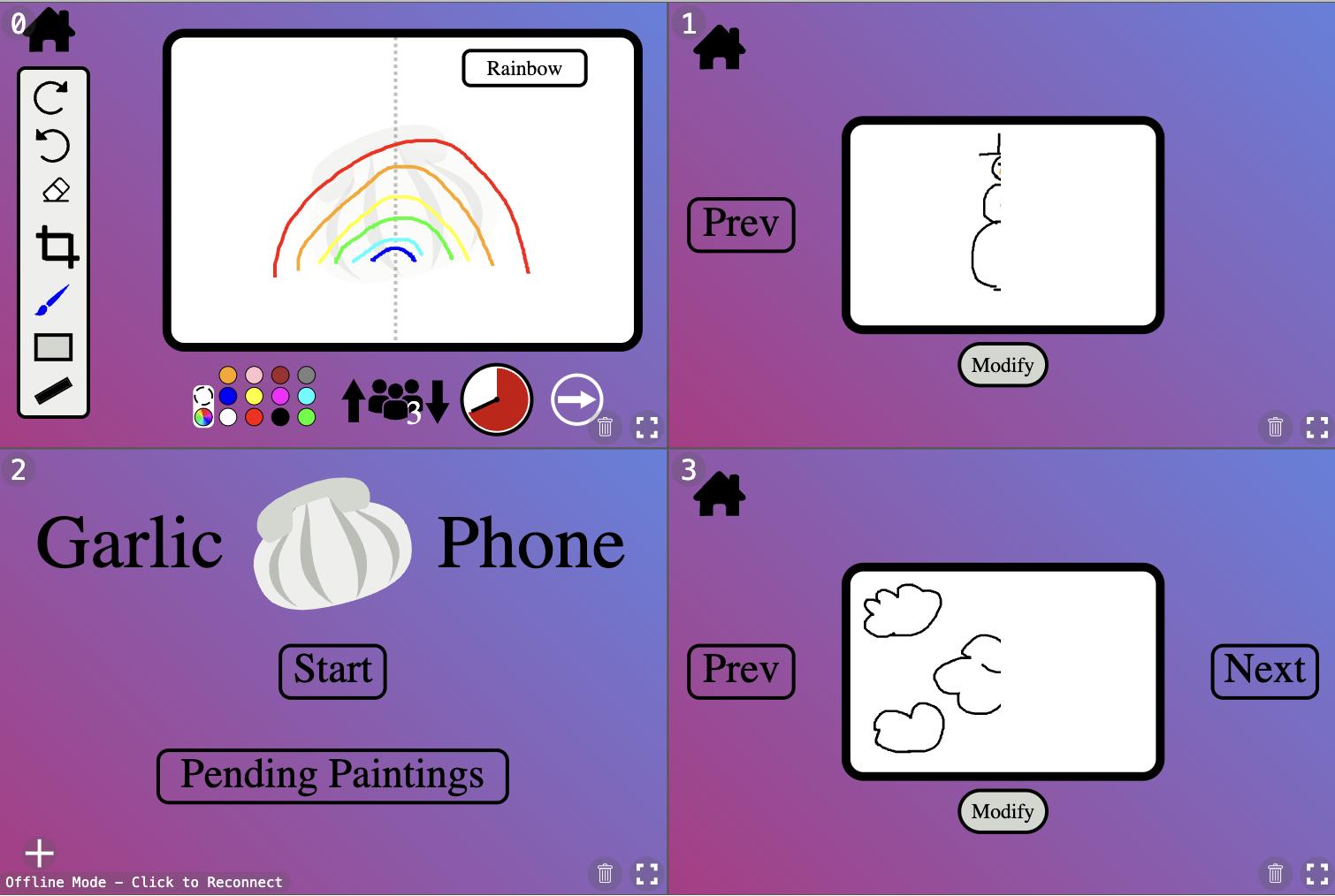}
    \caption{The multiplayer game ``Garlic Phone'', created by first-year CS students. In development mode, as displayed, multiple concurrent clients are simulated using a split-screen display.
    Players take turns completing the other half of drawings created by other players. Mouse data can be collected for PD symptoms such as tremors.}
    \label{Fig:GarlicPhone}
\end{figure}

Fig.~\ref{Fig:BrushstrokeJourney} and \ref{Fig:GarlicPhone} show two games 
completed by first-year design teams, both of which have a drawing theme.
Mouse movements can later be tracked to detect progression of PD symptoms such as 
tremors.

\section{Threats to Validity}
\label{Sec:ThreatsToValidity}

Due to the opt-in nature of the study, there is likely to be a self-selection
bias; that is, it is possible that students who were most engaged with the
course were the most likely to participate in the surveys and focus groups.
Incentives like a gift card draw for the surveys and free pizza for the focus
groups were used to try to mitigate this issue. Still, these results may not 
be representative of all students.

The two focus groups had a total of 18 participants, while the pre- and post-
project implementation surveys had 26 and 22 participants respectively. 
Therefore, these instruments may fail to provide a representative sample of the 
class.

\section{Discussion}
\label{Sec:Discussion}

From the average responses in Fig.~\ref{Fig:TopReasons} we see evidence that 
students were empathizing with their user,
avoiding \textbf{FMSub} and \textbf{FMHN}. 
Responses to these three questions indicate a 
willingness to prioritize creating a better project over following personal preferences,
supporting
comments from the focus groups which indicate that the PD problem engaged this cohort of students. 
The way they approached this, however, impacted their technology choices.
Teams producing multiplayer games were more likely to explain their choice as leading to a more engaging game (which was the goal of the project),
while teams producing single-player games were more likely to explain their choice as being a better fit for the DT problem, indicating a commitement to following the process.

The results in Fig.~\ref{Fig:ImportantFemale} and~\ref{Fig:ImportantMale} are consistent with research  \cite{hill2017gender} that has shown female-identifying students are more readily engaged when using technology to solve real problems instead of ``tinkering''. This suggests DT with problems like the PD problem can help reduce the gender imbalance in computing.

Fig.~\ref{Fig:UsefulTool}-\ref{Fig:PowerfulEnough} indicate that TEASync was 
generally viewed as useful, but somewhat difficult to learn and not necessarily capable of solving developer problems. 
In the focus groups, many reported already having committed
to a single-player game before TEASync was introduced.
In the future, the framework should be previewed early in the course to mitigate this problem.
But by using the TEASync framework to add a leaderboard to a successful single-player game, the Brushstroke Journey (Fig.~\ref{Fig:BrushstrokeJourney}) team epitomizes  appropriate use of technology---demonstrating both technical mastery and focus on the user in making design decisions.
Anecdotally, we can report a much higher level of achievement overall, relative to previous cohorts given different design problems. 

Different mitigation strategies help to avoid common problems learners
face with DT when they are starting out.
These are listed together with the aspects of course design which mitigated against them in 
 Fig.~\ref{Fig:FMMitigation}. These
include following good problem design properties (see 
Fig.~\ref{Fig:GoodProblemProperties}), course design (i.e., requiring traceability from user needs to design choices), as well as
tool design (i.e. the proposed worksheets). Many failure modes are 
mitigated through a combination thereof.
Note that worksheet design, including the Game Matrix, are included under ``Tool Design.''

\begin{figure}[]
    \centering
    \small
    \rowcolors{2}{gray!15}{white}
    \caption{Summary of Failure Mode Mitigation Strategies}
    \label{Fig:FMMitigation}
    \begin{tabular}{p{0.075\textwidth}ccc}
     \toprule
      \textbf{Failure Mode} & \textbf{Problem Design} & \textbf{Course Design} & \textbf{Tool Design}\\
        \midrule
        FMSub & \checkmark & & \\
        FMHN  & \checkmark & \checkmark & \checkmark \\
        FMSC  & & & \checkmark \\
        FMJ   & \checkmark & & \checkmark\\
        \bottomrule
    \end{tabular}
    \vspace{0mm}
    
    Legend: FMSub = Substitution; FMHN = Hammer-Nail Problem; FMSC = Idea Self-Censoring; FMJ = Jump to the First...
\end{figure}

\section{Conclusions}
\label{Sec:Conclusions}

A combination of a well-scaffolded method and a good problem
design, together with EDP and MDD tools providing an easy on-ramp for beginners and scope for advanced reasoning (i.e., about concurrency) are sufficient to ensure a successful first-year design course at scale,
with modest instructional resources. 
We believe our method is generalizable and effectively teaches the empathy skills and appropriate decision making needed to be successful software designers.

\subsection{Generalizability}
The worksheets were set up for the PD project and assumed that
the final output would be a game. However, generalizing this approach to 
other courses, such as a Capstone course is straightforward using the following steps:

\begin{enumerate}
    \item Update timetable based on course requirements. You can also
    adjust the number of DT updates, though we have found that four (due approximately every two weeks) is a good number to keep groups on track.
    \item Update research slides to your problem, or leave them blank if groups choose their own problem.
    \item The users' interest slides can be changed to another aspect,
    if it doesn't make sense in the project context. For very open-ended projects, this can be a set of related problems the groups are
    thinking of focusing on.
    \item The Game Matrix can be generalized as the ``Brainstorming Matrix'' or ``Idea Matrix''. 
    The rows and columns need to be independent aspects of the problem/solution.
    It is possible to ask design teams to populate the row/column headers themselves,
    but we strongly encourage instructional teams to give feedback and final approval 
    on the headers before the matrix is completed.
\end{enumerate}

\subsection{Future Work}
\label{SubSec:FutureWork}
One area of exploration is  co-design, in which users collaborate with designers to produce a design, and are considered part of the team \cite{sanders2008co}. 
The use of co-design aims to produce higher quality products, better alignment between the product's functionalities and the users' needs, and increased user satisfaction \cite{wintermans2017together}. 
We have run a small pilot with high school students and their grandparents with promising results.

It would be useful to establish metrics for evaluating design problems so that experiences in different institutions could be pooled, and the best problems identified.

In terms of tools, our slide-based worksheets approach is usable in its current 
form, but based on feedback and observation, lacks support for traceability and
searchability. Future work will seek to create a bespoke platform 
for collaborative DT which explicitly tracks traceability from the 
empathize stage through to prototyping and iteration. 

\section*{Acknowledgments}

We thank the students who shared their experiences via surveys and focus groups,
and final projects.

\bibliographystyle{IEEEtran}
\bibliography{bib}

\end{document}